\documentclass[doublecolumn]{IEEEtran}
\usepackage{amsmath,amsthm,amsfonts,amssymb,bm,mathrsfs}
\usepackage{subfigure}
\usepackage{graphicx,multicol}
\usepackage{algorithm}
\usepackage{algorithmic}
\usepackage{multirow,booktabs,threeparttable,array}
\usepackage{color,cite,url}

\usepackage{setspace}
\theoremstyle{remark}

\theoremstyle{plain}

\begin{document}
\title{Toward Native Artificial Intelligence in 6G Networks: System Design, Architectures, and Paradigms}

\author{
Jianjun Wu, Rongpeng Li, Xueli An, Chenghui Peng, Zhe Liu, Jon Crowcroft, and Honggang Zhang

}

\maketitle

\begin{abstract}
The mobile communication system has transformed to be the fundamental infrastructure to support digital demands from all industry sectors, and 6G is envisioned to go far beyond the communication-only purpose. There is coming to a consensus that 6G will treat Artificial Intelligence (AI) as the cornerstone and has a potential capability to  provide ``intelligence inclusion", which implies to enable the access of AI services at anytime and anywhere by anyone. Apparently, the intelligent inclusion vision produces far-reaching influence on the corresponding network architecture design in 6G and deserves a clean-slate rethink. In this article, we propose an end-to-end system architecture design scope for 6G, and talk about the necessity to incorporate an independent data plane and a novel intelligent plane with particular emphasis on end-to-end AI workflow orchestration, management and operation. We also highlight the advantages to provision converged connectivity and computing services at the network function plane. Benefiting from these approaches, we believe that 6G will turn to an ``everything as a service" (XaaS) platform with significantly enhanced business merits.  
\end{abstract}

\section{Introduction}
Mobile communication has stepped into a new and exciting era, where personal devices such as smartphones and tablets are becoming the primary computing platform for many applications. As these devices are used to facilitate our daily lives (from professional to leisure and education to entertainment), they simultaneously generate an unparalleled amount of personal and private data. Meanwhile, the 3rd Generation Partnership Project (3GPP) Release 15, the first full set of 5G specifications, was frozen in the middle of 2018, and the first commercial 5G launch happened in 2019 in Korea, signaling that 5G has been transformed from standardization to commercialization. Even though 5G itself still needs several releases for gradual enhancement, it is the time to begin to think, what will happen for beyond 5G. 

\begin{figure}
\includegraphics[width=9cm]{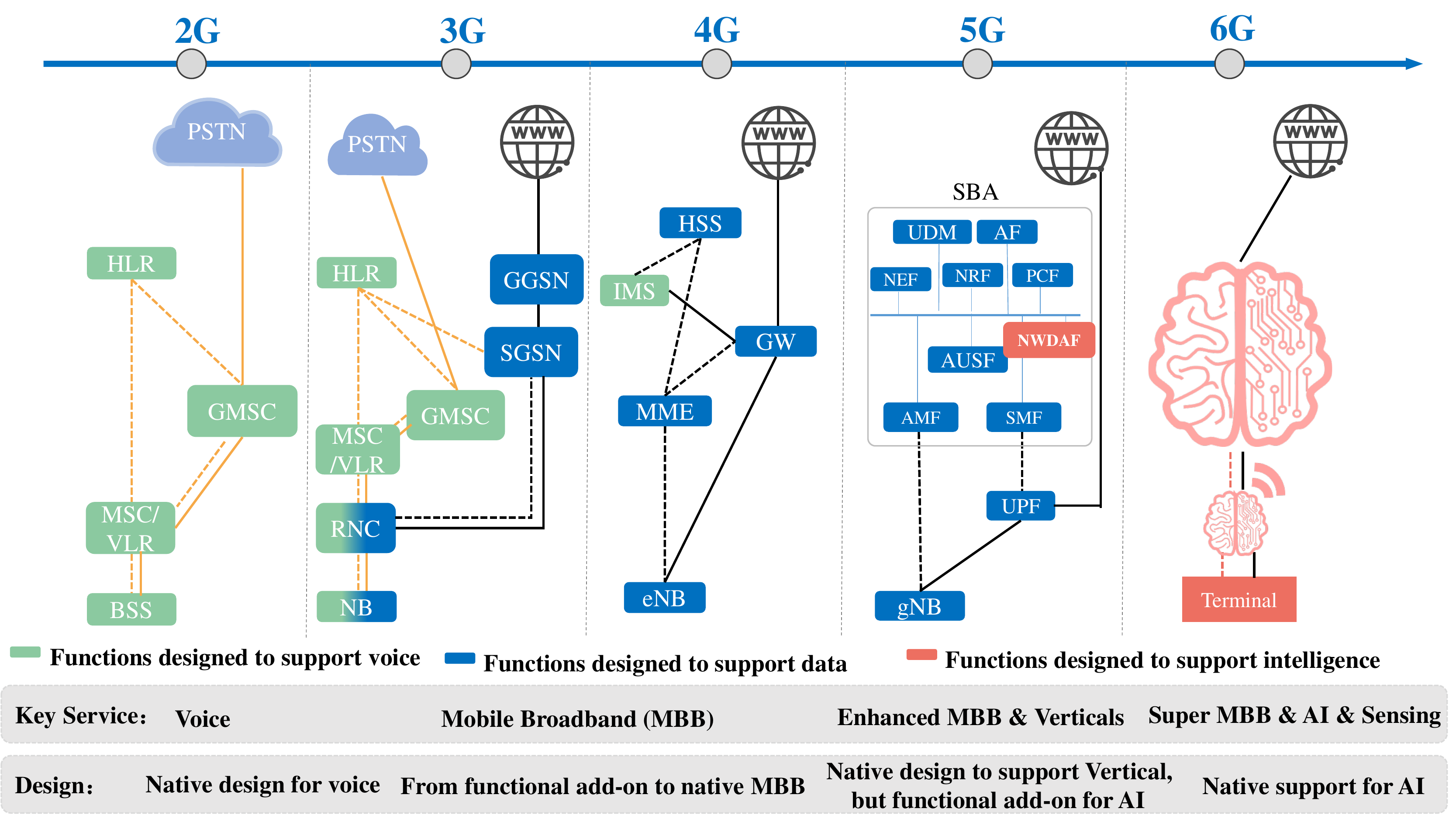}
\caption{Mobile communication system evolution trend.}
\label{fig:fig_history}
\vspace{-0.5cm}
\end{figure}

From the perspective of mobile network evolution history as shown in Figure~\ref{fig:fig_history}, the provided services intrinsically determinate the native architecture, such as a voice-centric network architecture in 2G, the incrementally functional add-on methods to offer data services upon voice services in 3G, and a data-native network architecture based on all-IP solutions emerged in 4G.
%
5G begins a strong focus to support the digital transformation of vertical industries and significantly improve the efficiency of the entire social-economic activities. Therefore, 5G adds killer capabilities like enhanced mobile broadband (eMBB), ultra reliable low latency (URLLC) and massive machine-type communication (mMTC) and establishes a native service based architecture (SBA) especially for the core networks (CN). Benefiting from the cloudification and virtualization technologies, the SBA in 5G is capable to dramatically reduce the standardization complexity especially on relevant network function, interface and procedure definitions for newly introduced services. Such features will be definitely carried on towards to the next generation. Therefore, at this point in time, it would be natural and essential to think and reflect, which new services will be provided by 6G and how to invoke the required network architecture design innovation.

On the other hand, artificial intelligence (AI) has experienced promising development in the past decades. The advanced progress in algorithms, platform as well as chipset technology lets AI receive the sustainable momentum and bring its impacts to many different industry sectors, which includes telecommunication sector as well. 5G has initiated to provide a network design suitable for AI services by introducing the Network Data Analytics Function (NWDAF)~\cite{3GPP501} to the 5G CN to implement AI-based network automation and optimize the related network functions (such as AI-based mobility management). Its base is to collect and analyze data from other 5G network elements to train AI models that can be used by network services. Meanwhile, similar mechanism like collecting and analyzing data based on the existing SON/MDT (Self-Organizing Networks and Minimization of Drive Tests) is adopted for 5G radio access networks (RAN).
Nevertheless, 5G retains the architecture scope of previous generations, and only focuses on how to use AI to optimize the network itself. What really matters is that the intelligent features can be utilized by the end users, which however is out of the scope of 5G architecture design. In this regard, the emergence of NWDAF is only the starting point to support AI.

Other than 3GPP, International Telecommunication Union (ITU) also initializes work on how to integrate AI framework within the mobile communication system. Their focus group has drafted several technical specifications for machine learning for future networks, including interfaces, network architectures, protocols, algorithms and data formats~\cite{itu}. Academic research also provides meaningful results to understand the relationship between mobile communication technology and AI. For instance, AI technology could be used to optimize and enhance network resource usage in order to achieve better performance. At the radio link layer, radio resource scheduling or link adaptation can be better implemented through radio channel prediction. Or at the network layer, AI is used to solve the problem of self-driving network operation and management or the twinning of future networks~\cite{li_intelligent_2017,Bega20}. On the other hand, it is also envisioned that enhanced network capability that could provide better support for AI services. AI would benefit from better connections with pre-processing computing power~\cite{li_collective_2020}. 

6G has been on its journey since 2020. Many leading companies and industry platforms have announced their vision on 6G. There is coming to a consensus that 6G will treat AI as the cornerstone. Finnish 6G initiatives has published a series of white papers~\cite{6gflag} to lay out their vision on 6G technologies, in which the authors also place their prediction on future trend, ``\textit{the evolution of telecom infrastructures towards 6G will consider highly distributed AI, moving the intelligence from the central cloud to edge computing resources}". In~\cite{Zhu20}, the authors introduced a set of principles for the intelligent edge for wireless communication with embedded machine learning capabilities. 
However, if we simply take the edge computing as it is, being designed orthogonal of the network architecture, it cannot provide native AI support for the whole mobile communication system, which is devastatingly essential in the near future. 

This article aims to use a clean slate approach to define 6G end-to-end (E2E) system architecture providing native support for intelligence inclusion, so as to better leverage infrastructure resources, facilitate data governance, system orchestration and management, as well as mobilize the enthusiasm of all stakeholders. Upon such design, 6G networks will support secured and high performance converged computing and communication, data governance framework, and collaborative ecosystem. Finally, we also provide some initial case studies for proof of concept.

 \begin{figure*}[h]
\includegraphics[width=\textwidth]{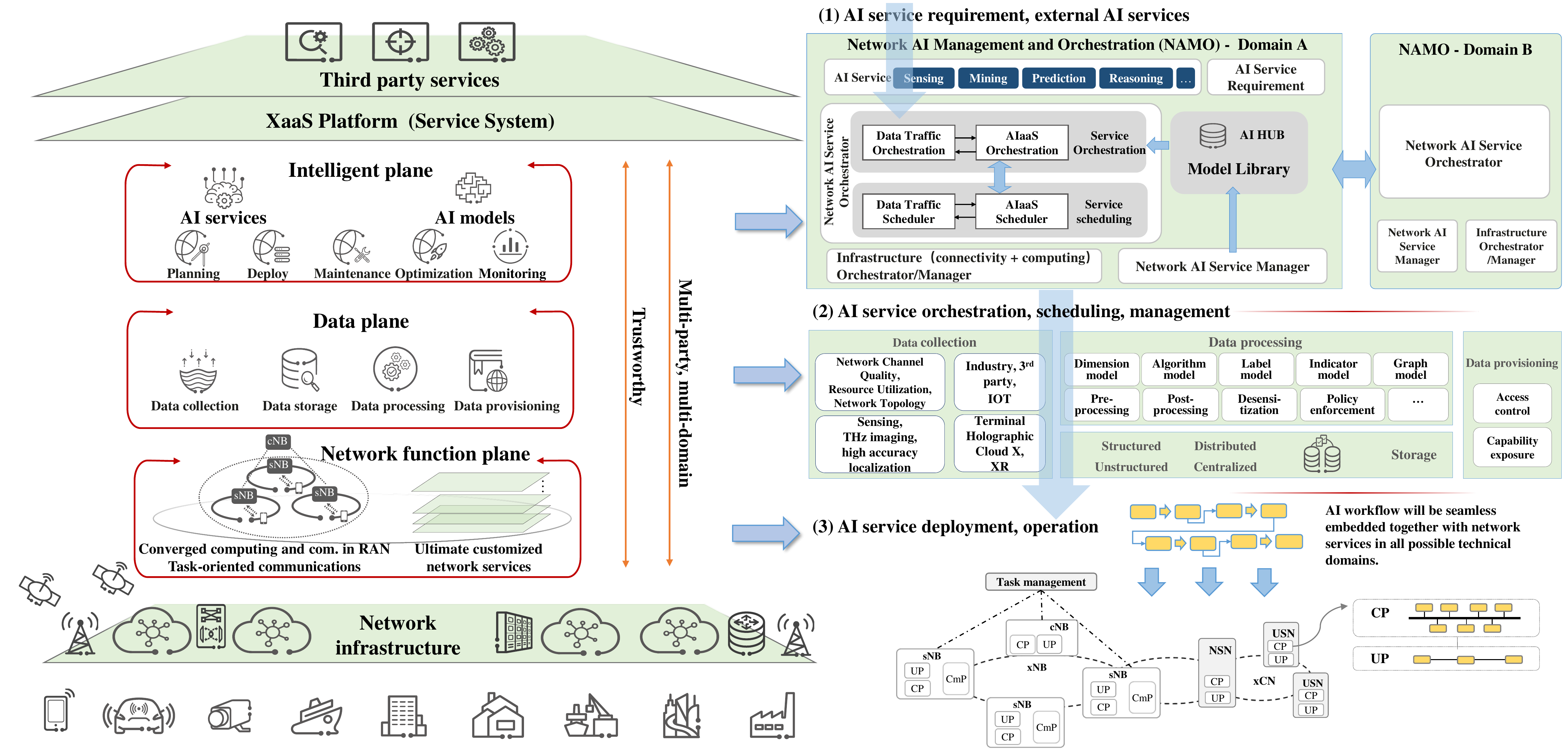}
\caption{6G system with high level architectures for intelligence inclusion.}
\label{fig:fig_arc}
\vspace{-0.5cm}
\end{figure*}

 \section{6G System Architecture Design Scope}
 ``Intelligence inclusion", which is completely different from the ``connectivity inclusion" in 5G and its precursors, is a high-level abstraction of our early vision of 6G to provide native support for AI services. Intelligence inclusion implies that 6G system should bring AI from the central cloud down to the mobile communication system and everyone can have the same capability to be able to access intelligent services anytime and anywhere. The merits of intelligence inclusion are two-folded. Firstly, running AI at any possible location in a distributed manner could further boost the learning efficiency and resolve critical security concerns. Secondly, AI could also help 6G system to realize enhanced network autonomous driving, which will further improve system performance (e.g., throughput, latency, spectrum efficiency as well energy efficiency) and user experience via online optimization. Hence, a true \textit{Pervasive Intelligent} mobile system, where communications and computing resources are deeply converged and corresponding instruments (e.g., algorithms, neuronal networks, solvers, databases, and APIs) are effectively integrated, will become an essential trait.
 However, such a vision does not come at no cost. For example, real-time learning and inference on ubiquitous connectivity and computing resources imposes stringent requirements (e.g., ultra-low latency and ultra-high data rates) on the mobile system and necessitate a fundamental rethink on the system architecture, so as to fully unleash the potential of the three major pillars (i.e., data, resource, and algorithms). Accordingly, we foresee that the architecture changes should incorporate at least the following methodologies: 
\begin{itemize}
  \item Infrastructure: 
  Shift from conventional connectivity-oriented service provisioning to handling computing-intensive tasks like AI. 
  \item Data governance: Handle complex data collections from different technical and business domains and provide data service which complies with security and privacy regulation. 
  \item Operation Administration and Maintenance: Provide E2E orchestration and management for AI workflows, which is seamlessly integrated with network resource orchestration and management.  
  \item Third-party AI services: Easy and flexible deployment deep in the mobile communication system.
\end{itemize}

%
Under this design methodology, we introduce the overall logical architecture for 6G especially as illustrated in Figure~\ref{fig:fig_arc}. In essence, there are several key features: (1) An independent data plane, (2) A novel intelligent plane, (3) Converged communication and computing service provisioning at network function plane, (4) XaaS (everything as a service) platform for a healthy and robust mobile ecosystem. Generally speaking, these features from different planes, which will be elaborated later, shall all consider the following two aspects.

(1) Security: Since 6G becomes the digital foundation of the entire society, it faces more serious security issues than before, especially when facing threats from advanced technologies like quantum computing. Therefore, traditional cryptography security based on computing complexity will become less qualified. In addition, the introduction of new technologies like AI inevitably bring more security risks and heavily require new mechanism to combat increasingly intelligent attacks. 

(2) Multi-party collaboration: The openness of the 6G ecosystem will attract different business entities (e.g., communication operators and vendors, end users, AI service providers) from both information and communication industry (ICT) and other vertical industries. Therefore, the system design with seamless multi-party support is a crucial feature. 
%


\section{Network Function Plane}
The network function plane of 6G system lays the very foundation to provide converged communication and computing services across different technical domains, including RAN, CN, and transport networks (TN). In particular, it is expected to support a large-scale distributed system with intelligence and sensing capability, so as to better reap gains from massive amounts of data generated from mobile communication system itself. 
Therefore, the network function plane should inherit the fundamental design principle like SBA and cloudification. Meanwhile, it needs to add new features on top. 

\subsection{Deep Converged Communication and Computing in RAN}
6G RAN is envisioned to embrace heterogeneous resources organized in ad hoc manner, which may cover terrestrial and non-terrestrial communication at the same time. Base stations (BSs) could be fixed, or mobile (e.g. cars, drones, or even satellites). Moreover, 6G RAN should be capable to support dynamic AI service deployment with critical performance requirements. 

Intelligence inclusive RAN is more than installing racks of servers at the edge location and local break out of the traffic for edge processing. Instead, it should provide deeply converged communication and computing capability. Therefore, a 6G BS (denoted as xNB for consistency with 5G) can be further divided into two parts: control base station (cNB) and service base station (sNB). The cNB provides control function (e.g., including connectivity and computing control) in a large area, while the sNB provides high-speed data transmission and computing services in a smaller area, e.g., cell level. The separation of the cNB and the sNB has the following advantages: 
(1) cNB can provide more efficient resource coordination and heterogeneous resource convergence control among sNBs and user equipment (UEs); 
(2) It can facilitate the deployment, since cNB can provide good coverage with lower frequency, while sNB can provide good throughput and energy-efficiency with higher frequency; 
(3) it can reduce overhead for common control purpose. 

Both cNB and sNB share some similarities with the 5G RAN like the control plane (CP) and user plane (UP) functions. For example, the control functions provided by the cNB include common and UE-specific computing resource control and radio resource control (such as Master/System Information Block (MIB/SIB), paging, and UE-specific radio resource control (RRC) signaling). Meanwhile, we argue that it is worthwhile to introduce an independent computing plane (CmP), which could be utilized to host certain services (e.g., AI or other third party applications) and other extended CN functions. As shown in Figure~\ref{fig:fig_arc}, the sNB is responsible to logically separate the CP, UP and CmP services.

When AI services are deployed at RAN level, they need to establish the connection between terminals and these functions directly in order to provide high-performance communication. This is essential especially for real-time AI services. Therefore, new interfaces between CmP and CP, and between CmP and UP shall be defined to support the dynamic service deployment, which will be further explored in our future work.

\subsection{Task-oriented Connectivity}
Conventional communications systems are connection-oriented, for which a typical service could be establishing connection between two specific terminals. Therefore, the communication source and destination are clearly defined by end users and the services that they intend to use or the other users that they plan to communicate with. In 6G, other than connectivity-oriented services, it should provide AI-based services, for instance predictive QoS for a car to perform full autonomous driving. In order to cater such services, connections are explicitly or implicitly established between many terminals and network equipment in a proactive or reactive manner and computation resource shall be scheduled as well. Hence it requires to perform coordination and communication among multi-terminals, multi-xNBs and multi-computing resources in the form of \textit{Tasks}. Typical scenarios of tasks could be the provisioning of AI services to enhance the reliability for field level communication in smart manufacturing scenario, or the adoption of AI services at the xNBs in a specific region for better resource utilization. In order to perform these tasks, \textit{task-oriented communication} happens by executing the very same actions (e.g., establishing connectivity, and computing resource allocation) across numerous distributed nodes with proper coordination. 
It poses new challenges to design the network function plane in 6G, and new functionalities may emerge in order to perform task management and control. For instance, as illustrated in Figure~\ref{fig:fig_arc}, task control plane (T-CP) and user plane (T-UP) could be specifically designed in cNB and sNB, respectively. 
Meanwhile, novel task protocol stack could be carried on 6G radio control signaling (e.g., RRC), or radio data protocol (e.g., PDCP or SDAP) to transmit task messages and implement task decomposition, distribution, activation, etc.
Task management entity might be designed outside of RAN, e.g., as a CN function.

\subsection{User-centric Network Service Provisioning}
In order to create a highly customized and secured network environment for users, user-centric concept shall be introduced to design the system architecture, so as to support users' participation in network service definition and operation, and provide users full control of data ownership. In order to fulfill this design, the network architecture shall be divided as user service node (USN) and network service node (NSN). For example, USN contains customized services and network policies at end user level to meet diversified service requirements in the 6G era. As a byproduct, it could further support to establish the user profiling as the reflection of physical entity.

\section {Data Governance Framework with Independent Data Plane}
Data is the key asset of the entire communication system in 6G. Since the involved data types and magnitude in 6G change dramatically from AI operation to management, from business to consumers, and from environment awareness to terminals, it becomes incentive to provide a unified and efficient data governance framework to efficiently collect, organize, desensitize, store and access the data, thus providing better support for other third-party data applications and satisfying data privacy protection requirements. Given the complications to resolve three critical concerns addressed below, such a framework can be more 
effectively implemented via an independent data plane as shown in Figure~\ref{fig:fig_arc}.  
\subsection{Data Collection and Ownership}
The 6G system will inherently generate a huge amount of diverse data from both technical and business domains. An independent data plane in 6G could contribute to organizing and managing data efficiently while also considering privacy protection.
(1) Infrastructure: all types of physical and virtual resources in the mobile communication system evolve to potential data sources. The data plane could help to accumulate the diverse data from the heterogeneous infrastructure.
%
(2) Operation and Business Supporting: This data plane could conveniently store and analyze useful information for network operation and management, including physical equipment status, system operation information, service provisioning information, etc. 
On the other hand, benefiting from all data related to the business logic (e.g., customer relationship, partner relationship management) and more importantly customer subscription related information (e.g., user personal data or enterprise level data), the data plane could grant customers (no matter subscriber type or enterprise type) the capabilities to have full control of their own data.   
(3) Vertical Industries: Similarly, vertical industries could find room to store their interested data related to use-case operation, admiration and maintenance. Also, the data could be safely stored and securely accessed via their own repositories.
(4) Terminal: The data plane could add special space for terminals to store computing and communication resource related data, service usage profile, and sensing knowledge. Notably, data collection shall comply with the regional or national data protection policies and regulations of the data source in terms of usage rights and obligations, e.g., GDPR (General Data Protection Regulations).   

\subsection{Data Processing}
Depending on the regulation and data usage policy, data processing functions could be executed on the data plane or it could be plugged in the other planes (e.g., if raw data cannot be exported to data plane, it shall be pre-processed in the data source). Nowadays, collecting and storing sensitive data is often accompanied with privacy risks and responsibilities of protecting the privacy embedded in the data. Therefore, it is essential to have data desensitization modules as the key data processing services in the data plane. Meanwhile, data policy enforcement modules in the data plane, which handle data according to the regulatory as well as non-regulatory rules (e.g., geographic restriction), will be executed by default to ensure the integrity and legitimate of data processing. Data plane contains data processing model libraries to support different data services. Moreover, it also exposes its capabilities to other planes and perform access control for entities to access and utilize its data services. 

\subsection{Data Storage and Provisioning}
On data plane, (raw or processed) data could be stored flexibly in a structured or unstructured format through centralized or fully distributed method. Uniformed interfaces should be further defined to facilitate internal customers of the mobile communication system, e.g., network service plane, and intelligent plane. Moreover, it could also be easily extended to provide data to external customers, e.g., vertical industry players, and enterprises.

\section{Intelligent Plane}
The objective of the intelligent plane is to build comprehensive AI platform capabilities in the mobile communication system. Network AI Management and Orchestration (NAMO), which is responsible for orchestrating, managing, and scheduling E2E network AI-related services and resources, is the essential design for intelligent plane. Basically, NAMO should answer the following questions: 
(1) How to leverage the 6G system capability and provision real-time and high reliable AI services?
(2) How to orchestrate and manage heterogeneous and distributed resources?  (3) How to define a universal and efficient mechanism to provide diversified AI services, such as perception, data mining, prediction, reasoning? (4) How to orchestrate seamless together with other network services? To answer the above raised questions, NAMO defines the following key processes.
\subsection{AI Service Orchestration}

AI service orchestration focuses on the parsing and orchestration of network AI services. An AI service is mapped to a logical AI workflow, which refers to a set of (independent) processing modules. Depending on specific AI services, logical AI workflows could consist of several modules that utilize different AI models, such as support-vector machine (SVM), recurrent neural network (RNN), long short-term memory (LSTM), and convolutional neural network (CNN). 

The logical AI workflow may obtain data from the data plane, and the entry module may further perform data processing in sequence. This processing may be serial processing or based on directed acyclic graph. In addition, logical AI workflow shall take QoS requirements of the related services into consideration. 
Logical AI workflows are mapped to physical resources during service deployment. Each processing module defined in the logic AI workflow needs to be assigned with specific parameters or hyperparameters, which are stored as models in the AI hub. If basic data services are involved, e.g., data collection and pre-processing, relevant functional modules shall be used from the data plane. 

It is worthwhile to note here that many deep learning technologies are extremely inefficient (e.g., in terms of energy use). There are many works towards optimizing neural networks by a certain extent of compressing (e.g., audio and image processing in smart phones). It is interesting and essential to understand the capability constraints and the feasibility of AI technologies before performing AI service orchestration. 

\subsection{Infrastructure Orchestration}

AI workflows need to be mapped to specific resources for execution, including relevant components (e.g., a graphical processing unit, and a network connection). This is implemented by the infrastructure orchestrator upon heterogeneous and distributed resources at large scale. 

\subsection{AI Service and Infrastructure Management}

As the names imply, the AI service manager focuses on the status maintenance and management of AI services, by issuing administration policies to add, modify, and delete logical and physical AI workflows. Similarly, the infrastructure manager is responsible for managing distributed communication and computing resources.
The foregoing NAMO function belongs to a logical function, and is specifically deployed and mapped to physical entities in different infrastructure layers, including xNBs, UEs, etc. 

 
\subsection{Standardization or Open Source}
NAMO should be a native multi-domain solution, where domains could be different regional networks of a mobile operator or networks from different mobile operators. Moreover, in order to enable easy deployment and attract external service providers, NAMO should be capable to coordinate resources through standardized interfaces and protocol. Put differently, AI services should be easily deployed in 6G (at the edge or centrally) and the intelligent plane should allow convenient service operation (e.g., deployment, update, migration, and termination). It is essential to understand which technical components fall into the scope of standardization (e.g., 3GPP), and which could rely on open source approach. For instance, typical AI workflow orchestration framework like Kubeflow and modelarts could be leveraged to accelerate the design of AI workflows. Continuous integration/continuous development (CI/CD) approach shall be adopted, especially to bring seamless service development and provisioning.

\section {XaaS Platform}
Comprehensive and collaborative ecosystem, where a significant amount of AI applications will be easily deployed, surely emerges as the key feature for 6G era. Apart from AI-enabled network functions such as AI-empowered channel prediction and AI-empowered mobility management, 6G could benefit third-party AI applications from commercial partners and consumers like autonomous driving and smart manufacturing, which require significant space and time flexibility in resource usage and share many similar capability components. Owing to the introduced data plane and intelligent plane, the development, deployment, and management efficiency of these AI applications are greatly optimized. As a result, 6G turns to a XaaS platform providing communication and computing infrastructure services at the infrastructure as a service (IaaS) layer, basic component services at the platform as a service (PaaS) layer, and AI application services as a service (SaaS) layer. Such a 6G XaaS platform also lays a solid foundation for business model innovation and reflects the benefits of intelligent inclusion.


\section{Concept Illustration via Use Cases}
In this part, we first provide an overview of recent research progress on multi-agent intelligent systems, and highlight the potential contributions of networks to AI. Afterwards, as a special case, we experiment in an internet of vehicles scenario and demonstrate the positive impact of intelligence inclusive communications on AI.
\subsection{Role of Networks in Multi-Agent Intelligent Systems}
\begin{table*}
\caption{A list of typical multi-agent learning algorithms.}
\label{tb:mas}
\begin{footnotesize}
\begin{tabular}{m{2.8cm}m{6.4cm}cccc}
\toprule
\multicolumn{1}{c}{\multirow{2}{*}{Algorithm}} & \multirow{2}{*}{Summary} & \multicolumn{4}{c}{Key Features}   \\
 &  & Type   & Environment & Execution & Training \\
\midrule
IRL & Independent reinforcement learning. & Value-based  & No & Decentralized & Decentralized \\
\hline
DGN (ICLR 2020)  & Graph convolutional reinforcement learning. & Policy gradient & Cooperative   & Centralized   & Centralized   \\
\hline 
RIAL (NeurIPS 2016)   & A recurrent network for parameter sharing to train independent learning agents and address partial observability.    & Value-based     & Cooperative      & Decentralized & Centralized   \\
\hline 
DIAL (NeurIPS 2016)  & Centralized gradient sharing; Variant of RIAL.  & Policy gradient & Cooperative  & Centralized   & Centralized   \\
\hline
MADDPG (AAMAS 2018)  & Each agent is capable to access (or at least model) the exact parameters of other agents.  & Policy gradient & Mixed  & Decentralized & Decentralized \\
\hline
VDN (AAMAS 2018)  & Training individual agents with a value-decomposition network architecture.  & Policy gradient & Mixed  & Decentralized & Centralized   \\
\hline
QMIX (ICML 2018)  & Decompose the joint-action value functions into per-agent monotonic ones.  & Value-based  & Mixed & Decentralized & Centralized   \\
\hline
COMA (AAAI 2018)  & Use a centralized critic and a counter-factual advantage function based on solving the multiagent credit assignment. & Policy gradient & Mixed  & Decentralized & Centralized   \\
\hline
QTRAN (PMLR 2019)  & Transforming the joint action-value function into an easily factorizable one. & Value-based  & Mixed & Decentralized & Centralized  \\
\hline
SIRL (TNNLS 2021)   & Apply stigmergy and federated learning to enhance the communications among agents.  & Policy Gradient & Cooperative    & Hybrid   & Hybrid \\
\bottomrule
\end{tabular}
\end{footnotesize}
\vspace{-0.3cm}
\end{table*}

Basically, in large-scale multi-agent intelligent systems, the complex relationship among agents usually causes great difficulty for policy learning. In many multi-agent systems, the interactions between agents often take place locally, which means that agents neither need to coordinate with all other agents nor need to coordinate with others all the time. Traditional methods attempt to use pre-defined rules to capture the interaction relationship between agents. However, these methods cannot be directly used in a large-scale environment due to the difficulty of transforming the complex interactions between agents into operation rules. Table~\ref{tb:mas} provides a list of some typical multi-agent machine learning algorithms. It can be observed that these state-of-the-art multi-agent learning algorithms share some similarities, i.e., one of the most important part is that these algorithms require a centralized or decentralized training or execution mechanism and have to interact among agents by networks. In this regard, we argue the following networking ingredients are essential and will lately show that the networking ingredients could provide superior performance than totally independent multi-agent learning.
\begin{itemize}
    \item We can model the relationship between agents by a complete graph and leverage graph neural networks (GNNs) to further utilize the embedded relationship. 
    Given the wide existence of topology in cellular networks, GNNs could be naturally enhanced in the intelligence plane with the explicit knowledge from the data plane. 
    \item Federated Learning enables mobile terminals to collaboratively learn a shared prediction model while keeping all the training data on device, decoupling the gradient update and the original data. 
    Such a federated learning framework, which could significantly enhance the privacy, cannot neglect the effect of networks. 
    \item Stigmergy aims to complete complex tasks by a small amount of indirect information exchange like the pheromone left by ants in their walking paths. 
    Similarly, digital pheromone is often leveraged in the stigmergy to act as the intermediate information. Accordingly, local information update among partially networked agents becomes necessary, so as to update the digital pheromone and reduce the behavioral localities \cite{xu_stigmergic_2021}.
\end{itemize}

\subsection{Results of Network-Assisted Multi-Agent Learning}

\begin{figure}
    \centering
    \includegraphics[width=0.95\linewidth]{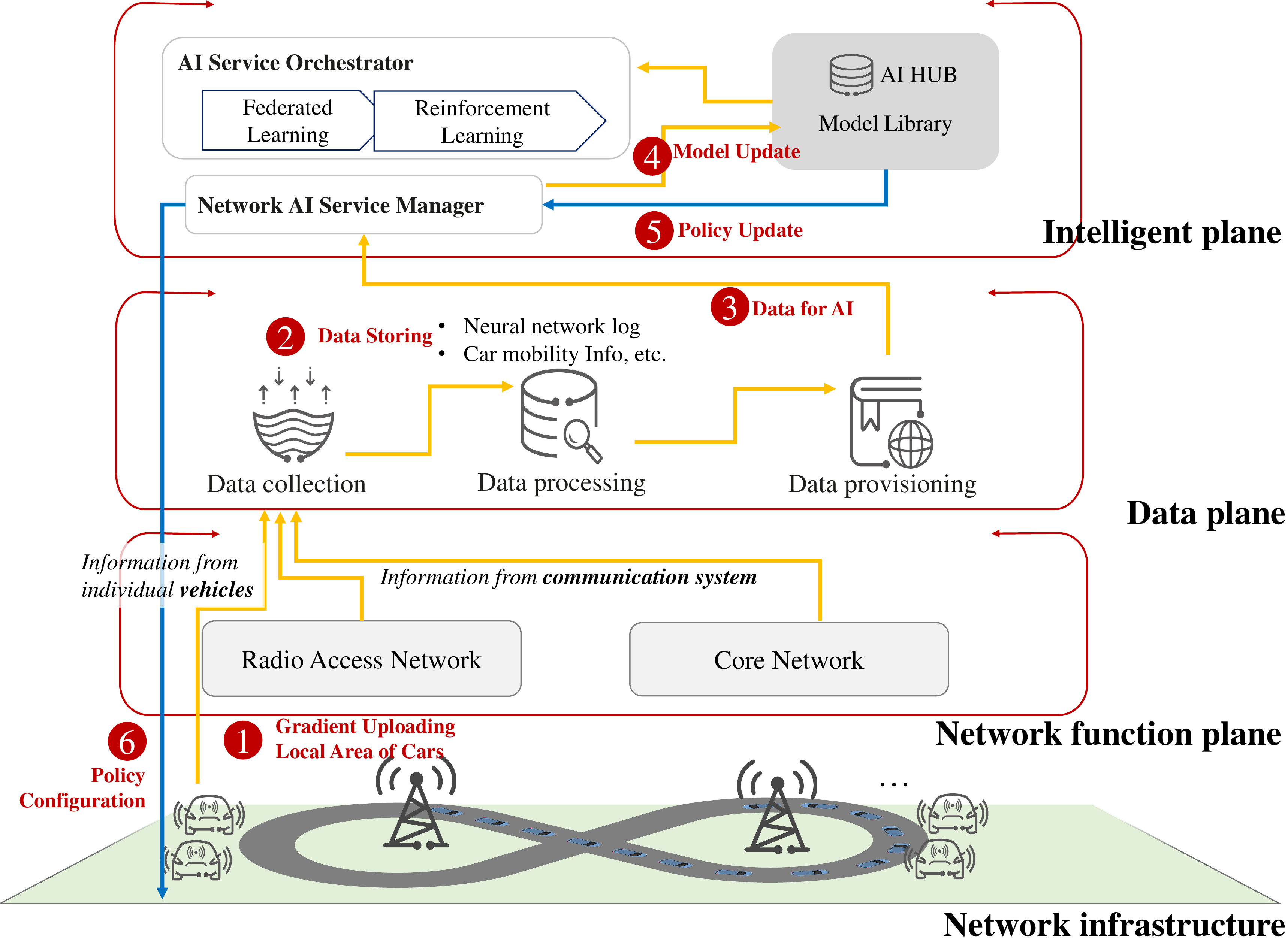} 
    \caption{The mixed-autonomy traffic control simulation scenario.}
    \label{fig:scenario}
    \vspace{-0.5cm}
\end{figure}

Next, we manifest the importance of networking in multi-agent systems. Due to the space limitation, we primarily consider the positive impact of network-assisted multi-agent learning via a mixed-autonomy traffic control scenario \cite{vinitsky_benchmarks_2018} as depicted in Figure \ref{fig:scenario}. As a representative showcase, there are totally 14 vehicles running circularly along a one-way lane that resembles a shape ``eight (i.e., 8)" and an intersection is located at the lane. Besides, each vehicle must adjust its acceleration to pass through this intersection in order to increase the average speed, but slamming on the brakes will be forced on vehicles that are about to crash, further terminating the current epoch, which is $150$ second. Furthermore in this article, the scenario is slightly modified by assigning the related local state of the entire environment to each vehicle, including the position and speed of its own, and the vehicle ahead and behind. Each vehicle collects its local state per $0.1$ second, and optimizes its acceleration accordingly by periodically computing the gradient from a $256$-minibatch. 

\begin{figure}
    \centering
    \includegraphics[width=0.95\linewidth]{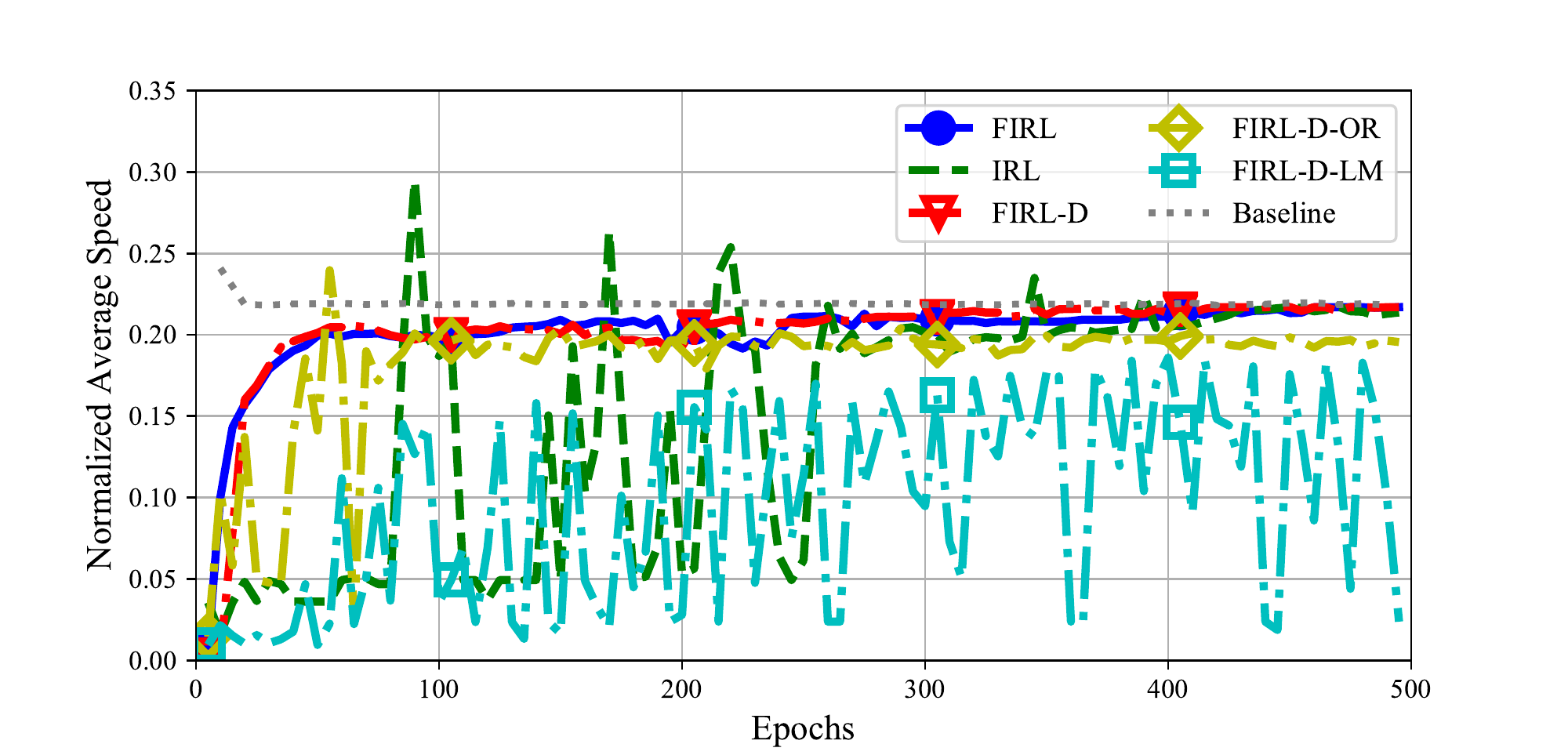} 
    \caption{The performance comparison of multi-agent learning with and without network-assistance.}
    \label{fig:performance}
\end{figure}

In order to testify the performance of multi-agent learning, these 14 vehicles can be categorized into two classes, namely, 7 cars underlying simulation of urban mobility (SUMO) and 7 independent reinforcement learning (IRL) or federated IRL (FIRL)-empowered cars, where FIRL implies the existence of network assistance and IRL means not. Besides, considering that SUMO is an open source, human being-level, highly portable and widely used traffic simulation package \cite{vinitsky_benchmarks_2018}, we treat 14 SUMO cars as the baseline. Figure~\ref{fig:performance} clearly verifies the advantage of FIRL over IRL and the gradual approaching of FIRL towards the baseline, and demonstrates the importance of network assistance.

Figure~\ref{fig:performance} also depicts three typical cloud-based AI (instead of inclusive AI) scenarios. Notably, the ``FIRL-D'' method simulates a direct latency increase by synchronously delaying the gradient uploading and the local neural network updating for all cars by 4 epochs and 2 epochs respectively. Furthermore, the ``FIRL-D-OR'' method checks the performance when the out-of-order among the neural network update packets occurs due to the multi-path and the fact that the latest neural network updated packet is possibly overridden by a previously delayed one. Meanwhile, ``FIRL-D-LM'' examines the case where the network quality is rather low and the uploading periodicity has to enlarge from $25.6$ second to $150$ second by locally merging 6 consecutive minibatchs. It can be observed that, synchronously increasing the network latency has trivial impact on the performance. However, the other cases leads to apparent performance degradation, which could better manifest the advantages to have the inclusive AI. 

\section{Conclusion and Future Works}
6G network architecture is envisioned to have inherent design that could provision inclusive intelligence for all types of end users. In order to realize this vision, this article provides an in-depth analysis from an E2E design perspective. For instance, how to achieve converged communication and computing capability at the deep edge via reconstructing the RAN, which is done by introducing computing capability in RAN that could be controlled and managed the same as connectivity services. Moreover, an independent data plane has been introduced to provide comprehensive data governance scheme, meanwhile an intelligent plane is established for E2E AI service orchestration and management. All these novel technical features could support flexible business models from XaaS platforms. To demonstrate the design scope described above, some initial case studies have been done via simulations, which indicates obvious advancement in some specific scenarios.

However, there are still a lot of research themes that need to be further explored. For example, whether this kind of architecture can adapt to all kinds of AI models, algorithms, etc. There could be new interfaces and protocols that need to be defined due to new introduced functionalities. In addition, it is also essential to consider how to build the right ecosystem that include all key players. Therefore, the architecture that inherently supports AI needs to be further improved in terms of technology and business ecosystem to truly enable the 6G intelligent inclusion vision.

\bibliographystyle{IEEEtran}

\bibliography{bib}







\end{document}